\documentclass[11pt]{llncs}
\usepackage[backend=bibtex8,giveninits=true,uniquename=init,maxnames=2,style=numeric-comp,doi=false]{biblatex}
\usepackage[british]{babel}
\usepackage{hyperref}
\usepackage{amsmath}
\usepackage{amssymb}
\usepackage{times}
\usepackage[babel=true]{csquotes}
\usepackage{booktabs} 
\usepackage{algorithm}
\usepackage{algpseudocode}
\usepackage{microtype}
\usepackage{xcolor}
\usepackage{nicefrac}

\newcommand{\uar}{\xleftarrow{\$}}

\newenvironment{myfont}{\fontfamily{phv}\selectfont}{\par}
\DeclareTextFontCommand{\textmyfont}{\myfont}
\newcommand{\Gen}{\textmyfont{KeyGen}}
\newcommand{\Enc}{\textmyfont{Enc}}
\newcommand{\Dec}{\textmyfont{Dec}}

\newcommand{\sk}{\textmyfont{sk}}

\AtBeginBibliography{\scriptsize}

\DeclareFieldFormat{eprint:iacr}{Cryptology ePrint Archive: \href{https://eprint.iacr.org/#1}{\scriptsize\texttt{#1}}}

\begin{document}
\title{Order-Preserving Encryption Using\\ Approximate Integer Common Divisors}
\author{James Dyer\inst{1} \and Martin Dyer\inst{2} \and Jie Xu\inst{2}}
\institute{{School of Computer Science, University of Manchester, UK.} \and {School of Computing, University of Leeds, UK.}\vspace{-3ex}}
\maketitle
\baselineskip 11.5pt

\begin{abstract}
We present a new, but simple, randomised \emph{order-preserving encryption} (OPE) scheme based on the \emph{general approximate common divisor problem} (GACDP). This appears to be the first OPE scheme to be based on a computational hardness primitive, rather than a security game. This scheme requires only $O(1)$ arithmetic operations for encryption and decryption. We show that the scheme has optimal information leakage under the assumption of uniformly distributed plaintexts, and we indicate that this property extends to some non-uniform distributions. We report on an extensive evaluation of our algorithms. The results clearly demonstrate highly favourable execution times in comparison with existing  OPE schemes.
\end{abstract}
\setlength{\textfloatsep}{3mm}
\setlength{\floatsep}{3mm}
\setlength{\intextsep}{3mm}

\setlength\abovedisplayskip{5pt plus 2pt minus 2pt}
\setlength\belowdisplayskip{5pt plus 2pt minus 2pt}
\setlength\abovedisplayshortskip{3pt plus 2pt minus 2pt}
\setlength\belowdisplayshortskip{3pt plus 2pt minus 2pt}
\section{Introduction}
\label{section:intro}
Outsourcing computation to the cloud has become increasingly important to business, government, and academia. However, in some circumstances, data on which those computations are performed may be sensitive. Therefore, outsourced computation proves problematic.

To address these problems, we require a means of secure computation in the cloud. One proposal, is that of \emph{homomorphic encryption}, where data is encrypted and computation is performed on the encrypted data \cite{rivest1978data}. The data is retrieved and decrypted. Because the encryption is homomorphic over the operations performed by the outsourced computation, the  decrypted result is the same as that computed on the unencrypted data.

\emph{Fully homomorphic encryption} has been proposed as a means of achieving this. However, as currently proposed, it is not practical. Therefore, we believe that \emph{somewhat homomorphic encryption}, which is homomorphic only for certain inputs or operations, is only of current practical interest.

For sorting and comparison of data we require an encryption scheme that supports homomorphic comparisons of ciphertexts. \emph{Order-preserving encryption} (OPE) is a recent field that supports just such a proposition. An OPE is defined as an encryption scheme where, for plaintexts $m_1$ and $m_2$ and corresponding ciphertexts $c_1$ and $c_2$,\footnote{This relationship is typically represented as $m_1 \leq m_2 \implies c_1 \leq c_2$. However, this seems to introduce an insecurity, by permitting an equality test for plaintexts using two comparisons.}
\setlength\abovedisplayskip{0pt}
\begin{align*}
    m_1 < m_2 \implies c_1 < c_2
\end{align*}
Our work presents an OPE scheme that is based on the \emph{general approximate common divisor problem} (GACDP) \cite{howgrave2001approx}, which is believed to be hard. Using this problem we have devised a system where encryption and decryption require $O(1)$ arithmetic operations.\vspace{-3ex}

\subsection{Notation}\vspace{-1ex}
$x \uar S$ represents a value $x$ chosen uniformly at random from the discrete set $S$.

$\Gen : \mathcal{S} \rightarrow \mathcal{K}$ denotes the key generation function operating on the security parameter space $\mathcal{S}$ and whose range is the secret key space $\mathcal{K}$.

$\Enc : \mathcal{M} \times \mathcal{K} \rightarrow \mathcal{C}$ denotes the symmetric encryption function operating on the plaintext space $\mathcal{M}$ and the secret key space $\mathcal{K}$ and whose range is the ciphertext space $\mathcal{C}$.

$\Dec : \mathcal{C} \times \mathcal{K} \rightarrow \mathcal{M}$ denotes the symmetric decryption function operating on the ciphertext space $\mathcal{C}$ and the secret key space $\mathcal{K}$ and whose range is the plaintext space $\mathcal{M}$.

$m,m_1,m_2,\ldots$ denote plaintext values. Similarly, $c,c_1,c_2,\ldots$ denote ciphertext values.

$[x,y]$ denotes the integers between $x$ and $y$ inclusive.

$[x,y)$ denotes $[x,y]\setminus\{y\}$.

$\mathbb{R}[x,y)$ denotes the real numbers in the interval $[x,y)$.\vspace{-3ex}

\subsection{Scenario}\vspace{-1ex}
Our OPE system is intended to be employed as part of a system for single-party secure computation in the cloud. In this system, a secure client encrypts data and then outsources computation on the encrypted data to the cloud. Then computation is performed homomorphically on the ciphertexts. The results of the computation are retrieved by the secure client and decrypted. We intend that our OPE scheme will support sorting and comparison of encrypted data.\vspace{-3ex}
\subsection{Formal Model of Scenario}\vspace{-1ex}
We have $n$ integer inputs, $m_1,m_2,\ldots,m_n$, where $m_i \in \mathcal{M}=[0,M]$ and $n \ll M$.\footnote{We must assume $n \ll M$ to avoid the ``sorting attack'' of Naveed et al. \cite{Naveed2015}} We wish to be able to compare and sort these inputs. A secure client $A$ selects an instance $\mathcal{E}_K$ of the OPE algorithm $\mathcal{E}$ using the secret parameter set $K$. $A$ encrypts the $n$ inputs by computing $c_i = \mathcal{E}_K(m_i)$, for $i \in [1,n]$. $A$ uploads $c_1, c_2, \ldots, c_n$ to the cloud computing environment. These encryptions do not all need to be uploaded at the same time but $n$ is a bound on the total number of inputs. The cloud environment conducts comparisons on the $c_i, i \in [1,n]$. Since $\mathcal{E}$ is an OPE, the $m_i$ will also be correctly sorted. $A$ can retrieve some or all of the $c_i$ from the cloud and decrypt each ciphertext $c_i$ by computing $m_i=\mathcal{E}_K^{-1}(c_i)$.

A snooper is only able to inspect $c_1, c_2, \ldots, c_n$ in the cloud environment. The snooper may compute additional functions on the $c_1, c_2, \ldots, c_n$ as part of a cryptanalytic attack, but cannot make new encryptions.\vspace{-2ex}

\subsection{Observations from Scenario}
\label{section:observ}
From our scenario we observe that we do not require public-key encryption as we do not intend another party to encrypt data. Symmetric encryption will suffice. Furthermore, there is no key escrow or distribution problem, as only ciphertexts are distributed to the cloud.

It is common in the literature \cite{bellare1998relations,bellare1997concrete} to refer to an encryption or decryption oracle in formal models of security. However, our scenario has no analogue of an oracle because another party has no way of encrypting or decrypting data without breaking the system. Any cryptological attacks will have to be performed on ciphertexts only. Therefore, we see \emph{chosen plaintext attacks} (CPA) and \emph{chosen ciphertext attacks} (CCA) as not relevant to our scenario. Indeed, it can be argued that any notion of indistinguishability under CPA is not relevant to OPE in practice (see section~\ref{section:models}). Various attempts have been made by Boldyreva and others \cite{Boldyreva2009,Boldyreva2011,teranishi2014oneway,xiao2012analysis} to provide such indistinguishability notions. However, the security models impose practically unrealistic restrictions on an adversary. See, for example, our discussion of IND-OCPA below (see section \ref{section:indocpa}). It should also be pointed out that satisfying an indistinguishability criterion does not guarantee that a cryptosystem is unbreakable, and neither does failure to satisfy it guarantee that the system is breakable.

We also note that a \emph{known plaintext attack} (KPA) is considered possible only by brute force, and not through being given a sample of pairs of plaintext and corresponding ciphertext.

Our notion of security requires only that determining the plaintext values is computationally infeasible within the lifetime of the outsourced computation. However, in some cases, we can show that the information leaked about the plaintexts is not significantly greater than is leaked by the total ordering revealed by the OPE.\vspace{-2ex}

\subsection{Related Work}
\label{section:relatedwork}
Prior to Boldyreva et al. \cite{Boldyreva2009}, OPE had been investigated by Agrawal et al. \cite{agrawal2004ope} and others (see \cite{agrawal2004ope} for earlier references). However, it wasn't until Boldyreva et al. that it was claimed that an OPE scheme was provably secure. Boldyreva et al.'s algorithm constructs a random order-preserving function by
mapping $M$ consecutive integers in a domain to integers in a much larger range $[1,N]$, by recursively dividing the range into $M$ monotonically increasing subranges. Each integer is assigned a pseudorandom value in its subrange. The algorithm recursively bisects the range, at each recursion sampling from the domain until it hits the input plaintext value. The algorithm is designed this way because Boldyreva et al. wish to sample uniformly from the range. This would require sampling from the negative hypergeometric distribution, for which no efficient exact algorithm is known. Therefore they sample the domain from the hypergeometric instead. As a result, each encryption requires at least $\log N$ recursions. Furthermore, so that a value can be decrypted, the pseudorandom values generated must be reconstructible. Therefore, for each instance of the algorithm, a plaintext will always encrypt to the same ciphertext. This implies that the encryption of low entropy data might be very easy to break by a ``guessing'' attack (see section \ref{section:results}). For our OPE scheme, multiple encryptions of a plaintext will produce differing ciphertexts. In \cite{Boldyreva2009}, the authors claim that $N=2M$, a claim repeated in \cite{Chenette2016}, although \cite{Boldyreva2011} suggests $N\geq 7M$. We use $N\geq M^2$ in our implementations of Boldyreva et al.'s algorithm, since this has the advantage that the scheme can be approximated closely by a much simplified computation, as we discuss in section \ref{section:approx}. The cost is only a doubling of the ciphertext size. However both \cite{Boldyreva2011,Boldyreva2009} take no account of $n$, the number of values to be encrypted. As in our scheme, the scheme should have $n\ll M$ to avoid the sorting attack of~\cite{Naveed2015}. If $c=f(m)$ is Boldyreva et al.'s OPE, it is straightforward to show that we can estimate $f^{-1}(c)$ by $\hat{m}=Mc/N$, with standard deviation approximately $\sqrt{2\hat{m}(1-\hat{m}/M)}$. For this reason, Boldyreva et al.'s scheme always leaks about half the plaintext bits.

Yum et al. \cite{Yum2012} extend Boldyreva et al.'s work to non-uniformly distributed plaintexts. This can improve the situation in the event that the client knows the distribution of plaintexts. This ``flattening'' idea already appears in \cite{agrawal2004ope}. In~\ref{section:flatten} we discuss a similar idea.

In \cite{Boldyreva2011}, Boldyreva et al. suggest an extension to their original scheme, modular order-preserving encryption (MOPE), by simply transforming the plaintext before encryption by adding a term modulo $M$. The idea is to cope with some of the problems discussed above, but any additional security arises only from this term being unknown. Note also that this construction again always produces the same ciphertext value for each plaintext.

Teranishi et al.\cite{teranishi2014oneway} devise a new OPE scheme that satisfies their own security model. However, their algorithms are less efficient, being linear in the size of the message space. Furthermore, like Boldyreva et al., a plaintext always encrypts to the same ciphertext value.

Krendelev et al. \cite{krendelev2014matrix} devise a an OPE scheme based on a coding of an integer as the real number $\sum_i b_i 2^{-i}$ where $b_i$ is the $i$th bit of the integer. The algorithm to encode the integer is $O(n)$ where $n$ is the number of bits in the integer. Using this encoding, they construct a matrix-based OPE scheme where a plaintext is encrypted as a tuple $(r,k,t)$. Each element of the tuple is the sum of elements from a matrix derived from the private key matrices $\sigma$ and $A$. Their algorithms are especially expensive, as they require computation of powers of the matrix $A$. Furthermore, each plaintext value always encrypts to the same ciphertext value.

Khadem et al. \cite{kadhem2010mv} propose a scheme to encrypt equal plaintext values to differing values. Their scheme is similar to Boldyreva et al. where a plaintext is mapped to a pseudorandom value in a subrange. However, this scheme relies on the domain being a set of consecutive integers for decryption. Our scheme allows for non-consecutive integers. This means that our scheme can support updates without worrying about overlapping ``buckets'' as Khadem et al.

Liu et al. \cite{liu2016new} addresses frequency of plaintext values by mapping the plaintext value to a value in an extended message space and splitting the message and ciphertext spaces nonlinearly. As in our scheme, decryption is a simple division. However, the ciphertext interval must first be located for a given ciphertext which is $\Omega(\log n)$ when $n$ is the total number of intervals.

Liu and Wang \cite{liu2012program} describe a system similar to ours where random ``noise'' is added to a linear transformation of the plaintext. However, in their examples, the parameters and noise used are real numbers. Unlike our work, the security of such a scheme is unclear.

In \cite{popa2013ideal}, Popa et al. discuss an interactive protocol for constructing a binary index of ciphertexts. Although this protocol guarantees ideal security, in that it only reveals the ordering, it is not an OPE. The ciphertexts do not preserve the ordering of the plaintexts, rather the protocol requires a secure client to decrypt the ciphertexts, compare the plaintexts, and return the ordering. It is essentially equivalent to sorting the plaintexts on the secure client and then encrypting them. Popa et al.'s protocol has a high communication cost: $\Omega(n\log n)$. This may be suitable for a database server where the comparisons may be made in a secure processing unit with fast bus communication. However, it is unsuitable for a large scale distributed system where the cost of communication will become prohibitive.  Kerschbaum and Schroepfer~\cite{kerschbaum2014oai} improved the communication cost of Popa et al.'s protocol to $\Omega(n)$ under the assumption that the input is random. However, this is still onerous for distributed systems. Kerschbaum~\cite{kerschbaum2015foe} further extends this protocol to hide the frequency of plaintexts. Boelter et al.\cite{boelter2016} extend Popa et al.'s idea by using ``garbled circuits'' to obfuscate comparisons. However, the circuits can only be used once, so their system is one-time use.

Also of note is \emph{order-revealing encryption} (ORE), a generalisation of OPE introduced by Boneh et al. \cite{Boneh2015}, that only reveals the order of ciphertexts. An ORE is a scheme $(C,E,D)$ where $C$ is a comparator function that takes two ciphertext inputs and outputs `$<$' or `$\geq$', and $E$ and $D$ are encryption and decryption functions. This attempts to replace the secure client's responsibility for plaintext comparisons in Popa's scheme with an exposed function acting on the ciphertexts.

Boneh et al.'s construction uses multilinear maps. However, as stated in Chenette et al. \cite{Chenette2016}, ``The main drawback of the Boneh et al. ORE construction is that it relies on complicated tools and strong assumptions on these tools, and as such, is currently impractical to implement''.

Chenette et al. offer a more practical construction, with weaker claims to provable security. However, since it encrypts the plaintexts bit-wise,  it requires a number of applications of a pseudorandom function $f$ linear in the bit size of the plaintext to encrypt an integer. The security and efficiency of this scheme depends on which pseudorandom function $f$ is chosen.

Lewi et al. \cite{lewi2016ore} devise an ORE scheme where there are two modes of encryption: left and right. The left encryption consists of a permutation of the domain and a key generated by hashing the permuted plaintext value. The right ciphertext consists of encryptions of the comparison with every other value in the domain. It is a tuple of size $d+1$ where $d$ is the size of the domain. Lewi et al. then extend this scheme to domains of size $d^n$. This results in right ciphertext tuples of size $dn+1$. Our experimental results compare favourably with theirs, largely because the ciphertext sizes of Lewi et al.'s scheme are much larger.

The security of these ORE schemes is proven under a scenario similar to IND-OCPA \cite{Boldyreva2009} (see section \ref{section:indocpa}). However, under realistic assumptions on what an adversary might do, these ORE schemes seem to have little security advantage over OPE schemes. For example, in $O(n\log n)$ comparisons an adversary can obtain a total ordering of the ciphertexts, and, hence the total ordering of the plaintexts. A disadvantage of ORE schemes are that they permit an equality test on ciphertexts~\cite[p.2]{Boneh2015} by using two comparisons. This could be used to aid a guessing attack on low-entropy plaintexts, e.g.~\cite{Naveed2015}. A randomised OPE scheme, like ours, does not permit this. On the other hand, the information leakage of the ORE schemes so far proposed appears to be near-optimal.\vspace{-3ex}

\subsection{Road Map}\vspace{-1ex}
\label{section:roadmap}
In section \ref{section:ope}, we present our OPE scheme. In section \ref{section:genericapprox}, we provide the generic version of Boldyreva et al.'s algorithm and the Beta distribution approximation used in our experiments. In section \ref{section:results}, we discuss the results of experiments on our OPE scheme. Finally, in section \ref{section:conc} we conclude the paper.\vspace{-1ex}

\section{An OPE scheme using Approximate Common Divisors}\vspace{-2ex}
\label{section:ope}
Our OPE scheme is the symmetric encryption system (\Gen, \Enc, \Dec ). The message space, $\mathcal{M}$, is $[0,M]$, and the ciphertext space, $\mathcal{C}$, is $[0,N]$, where $N>M$. We have plaintexts $m_i \in \mathcal{M}, i \in [1,n]$ such that $0<m_1\leq m_2\leq\cdots\leq m_n\leq M$.\vspace{-3ex}
\subsection{Key Generation}
Both the security parameter space $\mathcal{S}$ and the secret key space $\mathcal{K}$ are the set of positive integers. Given a security parameter $\lambda\in \mathcal{S}$, with $\lambda>\nicefrac{8}{3}\lg M$, $\Gen$ randomly chooses an integer $k\in[2^\lambda,2^{\lambda+1})$ as the secret key, \sk. So $k$ is a $(\lambda+1)$-bit integer such that $k>M^{\nicefrac{8}{3}}$ (see section \ref{section:security}). Note that $k$ does not necessarily need to be prime.\vspace{-2ex}
\subsection{Encryption}
To encrypt $m_i \in \mathcal{M}$, we compute,
\begin{equation*}
 c_i=\Enc(m_i,\sk)=m_i k + r_i,\vspace{-1ex}
\end{equation*}
where $r_i\uar(k^{\nicefrac{3}{4}},k-k^{\nicefrac{3}{4}})$.\vspace{-2ex}
\subsection{Decryption}
To decrypt $c_i \in \mathcal{C}$, we compute,
\begin{equation*}
  m_i=\Dec(c_i,\sk)=\lfloor c_i/k\rfloor.\vspace{-3ex}
\end{equation*}
\subsection{Order-preserving property}\vspace{-1ex}
 If $m>m'$, then $c\geq c'$ provided $mk+r>m'k+r'$, if $k(m-m')>(r'-r)$, which follows, since the lhs is at least $k$, and the rhs is less than $(k-1)$. If $m'=m$, then the order of the encryptions is random, since $\Pr(r'> r)\approx\frac12-1/k\approx \frac12$.\vspace{-2ex}
\subsection{Security of the Scheme}
\label{section:security}
Security of our scheme is given by the \emph{general approximate common divisor problem} (GACDP), which is believed to be hard. It can be formulated \cite{cohn2012approx,chen2012faster} as:
\begin{definition}[General approximate common divisor problem] Suppose we have $n$ integer inputs $c_i$ of the form $c_i=km_i + r_i$, $i\in[1,n]$, where $k$ is an unknown constant integer and $m_i$ and $r_i$ are unknown integers. We have a bound $B$ such that $|r_i| < B$ for all $i$. Under what conditions on $m_i$ and $r_i$, and the bound $B$, can an algorithm be found that can uniquely determine $k$ in a time which is polynomial in the total bit length of the numbers involved?
\end{definition}

GACDP and \emph{partial approximate common divisor problem} (PACDP), its close relative, are used as the basis of several cryptosystems, e.g.~\cite{dyer2017full,vandijk2010fully,coron2011fully}. Solving the GACDP is clearly equivalent to breaking our system. To make the GACDP instances hard, we need $k\gg M$ (see below). Furthermore, we need the $m_i$ to have sufficient entropy to negate a simple ``guessing'' attack~\cite{massey1994guessing}. However, note that the model in~\cite{massey1994guessing} assumes that we are able to verify when a guess is correct, which does not seem to be the case here.
Although our scenario does not permit it, even if we knew a plaintext,\,ciphertext pair $(m,c)$, it would not allow us to break the system, since $c/m=k+r/m\in [k,k+k/M]$, which is a large interval since $k\gg M$. A number $n$ of such pairs would give more information, but it still does not seem straightforward to estimate $k$ closer than $\Omega\big(k/(M\sqrt{n})\big)$. Thus the system has some resistance to KPA, even though this form of attack is excluded by our model of single-party secure computation.


Howgrave-Graham~\cite{howgrave2001approx} studied two attacks against GACD, to find divisors $d$ of $a_0+x_0$ and $b_0+y_0$, given inputs $a_0,b_0$ of similar size, with $a_0<b_0$. The quantities $x_0,y_0$ are the ``offsets''. The better attack in~\cite{howgrave2001approx}, \textsf{GACD\_L}, succeeds when $|x_0|,|y_0|< X=b_0^{\beta_0}$,  and the divisor $d\geq b_0^{\alpha_0}$ and
\[\beta_0= 1-\frac{1}{2}\alpha_0 - \sqrt{1-\alpha_0-\frac{1}{2}\alpha_0^2} - \epsilon.\]
where $\epsilon>0$ is a (small) constant, such that $1/\epsilon$ governs the number of possible divisors which may be output. We will take $\epsilon=0$. This is the worst case for Howgrave-Graham's algorithm, since there is no bound on the number of divisors which might be output.

Note that $\beta_0<\alpha_0$, since otherwise $\sqrt{1-\alpha_0-\frac{1}{2}\alpha_0^2}\leq 1-\frac{3}{2}\alpha_0$. This can only be satisfied if $\alpha_0\leq\frac{2}{3}$. But then squaring both sizes of the inequality implies $\alpha_0\geq\frac{8}{11}>\frac{2}{3}$, contradicting $\alpha_0\leq \frac{2}{3}$.

Suppose we take $\alpha_0=\frac{8}{11}$. Then, to foil this attack, we require $\beta_0\geq\frac{6}{11}$. For our system we have, $b_0-a_0=\max m_i-\min m_i=M$.\footnote{Note this is our $M$, not Howgrave-Graham's.} To ensure that the common divisor $k$ will not be found we require $b_0^{\alpha_0}\geq  k$, so we will take $k=b_0^{8/11}$. Since $b_0 \sim Mk$, this then implies $b_0=M^{11/3}$. Thus the ciphertexts will then have about 11/3 times as many bits as the plaintexts. Now \textsf{GACD\_L} could only succeed for offsets less than $b_0^{\beta_0}=b_0^{6/11}=k^{3/4}$. Thus, we choose our random offsets in the range {$(k^{3/4},\,k-k^{3/4})$}.

Cohn and Heninger~\cite{cohn2012approx} give an extension of Howgrave-Graham's algorithm to find the approximate divisor of $m$ integers, where $m>2$. Unfortunately, their algorithm is exponential in $m$ {in the worst case, though they say that it behaves better in practice. On the other hand,~\cite[Appendix~A]{cryptoeprint:2011:436} claims that Cohn and Heninger's algorithm is worse than brute force in some cases. In our case, the calculations in~\cite{cohn2012approx} do not seem to imply better bounds than those derived above.

We note also that the attack of \cite{chen2012faster} is not relevant to our system, since it requires smaller offsets, of size $O(\sqrt{k}$), than those we use.

For a survey and evaluation of the above and other attacks on GACD, see~\cite{galbraith_gebregiyorgis_murphy_2016}.\vspace{-2ex}

\subsection{Security Models}
\label{section:models}
\subsubsection{One-Wayness.}
\label{section:oneway}
The one-wayness of the function $c(m)=km+r$ used by the scheme clearly follows from the assumed hardness of the GACD problem, since we avoid the known polynomial-time solvable cases.\vspace{-2ex}

\subsubsection{IND-OCPA.}
\label{section:indocpa}
The model in~\cite[p.6]{Boldyreva2009} and \cite[p.20]{lewi2016ore} is as follows: given two equal-length sequences of plaintexts $(m_0^1 \ldots m_0^q)$ and $(m_1^1 \ldots m_1^q)$, where the $m_b^j$ ($b\in[0,1],j\in[1,q]$) are distinct,\footnote{~\cite[p.6]{Boldyreva2009} and \cite[p.20]{lewi2016ore} do not clearly state this assumption but it appears that all plaintext values used must be distinct. This assumption clearly does not weaken the model.} an adversary is allowed to present two plaintexts to a \emph{left-or-right oracle} \cite{bellare1997concrete}, $\mathcal{LR}^{(m_0,m_1,b)}$, which returns the encryption of $m_b$. The adversary is only allowed to make queries to the oracle which satisfy $m_0^i < m_0^j$ iff $m_1^i < m_1^j$ for $1 \leq i,j \leq q$. The adversary wins if it can distinguish the left and right orderings with probability significantly better than $\nicefrac12$.

However, Boldyreva et al. \cite[p.5]{Boldyreva2009} note, 
concerning chosen plaintext attacks: ``in the symmetric-key setting a real-life adversary cannot simply encrypt messages itself, so such an attack is unlikely to be feasible''. Further, they prove that no OPE scheme with a polynomial size message space can satisfy IND-OCPA. Lewi et al. \cite{lewi2016ore} strengthen this result under certain assumptions.

The IND-OCPA model seems inherently rather impractical, since an adversary with an encryption oracle could decrypt any ciphertext using $\lg M$ comparisons, where $M$ is the size of the message space. Furthermore, Xiao and Yen \cite{xiao2012} construct an OPE for the domain [1,2] and prove that it is IND-OCPA secure. However, this system is trivially breakable using a ``sorting'' attack \cite{Naveed2015}. For these reasons, we do not consider security models assuming CPA to be relevant to OPE. 
\vspace{-2ex}

\subsubsection{Window One-Wayness.}
\label{section:window}
We may further analyse our scheme under the same model as in~\cite{Boldyreva2011}, which was called \emph{window one-wayness}. The scenario is as follows. An adversary is given the encryptions $c_1\leq c_2\leq \cdots\leq c_n$ of a sample of $n$ plaintexts $m_1\leq m_2\leq \ldots\leq m_n$, chosen uniformly and independently at random from the plaintext space $[0,M)$. The adversary is also given the encryption $c$ of a challenge plaintext $m$, and must return an estimate $\hat{m}$ of $m$ and a bound $r$, such that $m\in(\hat{m}-r,\hat{m}+r)$ with probability greater than $\nicefrac12$, say.  How small can $r$ be so that the adversary can meet the challenge?

This model seems eminently reasonable, except for the assumption that the plaintexts are distributed uniformly. However, as we show in section \ref{section:flatten}, this assumption can be weakened in some cases for our scheme.

Since the $m_i$ are chosen uniformly at random, a random ciphertext satisfies, for $\mathbf{c}\in[0,kM)$,
\[\Pr(\mathbf{c}=c)=\Pr(k\mathbf{m}+\mathbf{r}=km+r)=\Pr(\mathbf{m}=m)\Pr(\mathbf{r}=r)=\frac{1}{M}\frac{1}{k}
=\frac{1}{Mk},\]
where $\mathbf{m}\uar[0,M)$, $\mathbf{r}\uar[0,k)$. Thus $\mathbf{c}$ is uniform on $[0,kM)$. Note that this is only approximately true, since we choose $\mathbf{r}$ uniformly from $[k^{\nicefrac{3}{4}},k-k^{\nicefrac{3}{4}}]$. However, the total variation distance between these distributions is $2Mk^{\nicefrac{3}{4}}/Mk=2/k^{\nicefrac{1}{4}}$. The difference between probabilities calculated using the two distributions is negligible, so we will assume the uniform distribution.

By assumption, the adversary cannot determine $k$ by any polynomial time computation. So the adversary can only estimate $k$ from the sample. Now, in a uniformly chosen sample $c_1\leq c_2\leq \cdots\leq c_n$ from $[0,kM)$, the sample maximum $c_n$ is a sufficient statistic for the range $kM$, so all information about $k$ is captured by $c_n$. So we may estimate $k$ by $\hat{k}=c_n/M$. This is the maximum likelihood estimate, and is consistent but not unbiased. The minimum variance unbiased estimate is $(n+1)\hat{k}/n$, but using this does not improve the analysis, since the bias $k/(n+1)$ is of the same order as the estimation error, as we now prove.
For any $0\leq \varepsilon\leq 1$,
\begin{align*}
\Pr\big(\hat{k}\in k(1\pm\varepsilon)\big)&\leq\Pr\big(c_n\geq kM(1-\varepsilon)\big)
\\
& = 1-(1-\varepsilon)^n\ \  \left\{\begin{array}{ll}
\ \leq n\varepsilon <\nicefrac12 &\ \hbox{if}\ \varepsilon< 1/(2n)  ;\\
\ \geq 1-e^{-n\varepsilon}\geq\nicefrac12  &\ \hbox{if}\ \varepsilon\geq\ln 2/n.
  \end{array}
\right.
\end{align*}
Now, if $c=mk+r$, we can estimate $m$ by $\hat{m}= c/\hat{k}\approx mk/\hat{k}$. Then
\[ \Pr\big(m\in\hat{m}(1\pm\varepsilon)\big)\approx\Pr\big(m\in mk/\hat{k}(1\pm\varepsilon)\big)
= \Pr\big(\hat{k}\in k(1\pm\varepsilon)\big) <\nicefrac12,\]
if $\varepsilon< 1/(2n)$. Thus, if $r\leq m/2n$, $\Pr(m\in\hat{m}\pm r)<\nicefrac12$. Similarly, if $r\geq m\lg 2/n$, $\Pr(m\in\hat{m}\pm r)\geq \nicefrac12$.  Thus the adversary cannot succeed if $r\leq m/2n$, but can if $r\geq m\lg 2/n$.

It follows that only $\lg m - \lg(m/n)+O(1)=\lg n +O(1)$ bits of $m$ are leaked by the system. However, $\lg n$ bits are leaked by inserting $c$ into the sequence $c_1\leq c_2\leq \cdots\leq c_n$, so the leakage is close to minimal. By contrast the scheme of~\cite{Boldyreva2009} leaks $\nicefrac12\lg m +O(1)$ bits, independently of $n$. Therefore, by this criterion, the scheme given here is superior to that of~\cite{Boldyreva2009} for all $n\ll\sqrt{M}$. Note that we have not assumed that $m$ is chosen uniformly from $[0,M)$, but the leakage of the random sequence $c_1\leq c_2\leq \cdots\leq c_n$ is clearly $n\lg n +O(n)$ of the $M\lg M$ plaintext bits. This reveals little more than the $n\lg n$ bits revealed by the known order $m_1\leq m_2\leq \cdots\leq m_n$.\vspace{-2ex}

\subsection{Further Observations}\vspace{-1ex}
This scheme can be used in conjunction with any other OPE method, i.e. any unknown increasing function $f(m)$ of $m$. We might consider any integer-valued increasing function, e.g. a polynomial function of $m$, or Boldyreva et al.'s scheme. If $f(m)$ is  this function, then we encrypt $m$ by $c=f(m)k+r$, where $r\uar(k^{3/4},\,k-k^{3/4})$,  and decrypt by $m=f^{-1}\big(\lfloor c/k\rfloor\big)$. 
The disadvantage is that the ciphertext size will increase.

If $f(m)$ is an unknown polynomial function, we solve a polynomial equation to decrypt. The advantage over straight GACD is that, even if we can break the GACD instance, we still have to solve an unknown polynomial equation to break the system. For example, suppose we use the linear polynomial $f(m)=a_1(m+a_0)+s$, where $s\uar [0,a_0]$ is random noise. But this gives $c=a_1k(m+a_0)+(ks+r)$, which is our OPE system with a deterministic linear monic polynomial $f(m)\gets m+a_0$, $k\gets a_1k$ and $r\gets ks+r\uar[0,a_1k)$, so $f(m)$ contains a single unknown parameter, $a_0$. More generally, we need only consider monic polynomials, for the same reason.

If $c=f(m)$ is Boldyreva et al.'s OPE, we can invert $f$ only with error $O(\sqrt{m})$. Therefore a hybrid scheme offers greater security than either alone.\vspace{-2ex}

\subsubsection{Flattening.}\label{section:flatten} Another use of such a transformation is when the distribution function $F(m)$ of the plaintexts is known, or can be reasonably estimated. Then the distribution of the plaintexts can be ``flattened'' to an approximate uniform distribution on a larger set $[0,N)$, where $N\gg M$. Thus, suppose the distribution function $F(m)$ ($M\in [0,M)$) is known, and can be computed efficiently for given $m$. Further, we assume that $\Pr(\mathbf{m}=m)\geq 1/N$, so $F$ is strictly increasing. This assumption is weak, since the probability that $\mathbf{m}$ is chosen to be an $m$ with too small probability is at most $M/N$, which we assume to be negligible.

We interpolate the distribution function linearly on the real interval $\mathbb{R}[0,M)$, by $F(x)=(1-u)F(m)+uF(m+1)$ for $x=(1-u)m+u(m+1)$, where $u\in\mathbb{R}[0,1)$. Then we will transform $m\in[0,M)$ randomly by taking
$\tilde{m}=NF(x)$ where $u$ is chosen randomly from the continuous uniform distribution on $\mathbb{R}[0,1)$. It follows that $\tilde{m}$ is uniform on $\mathbb{R}[0,N)$, since $F$ is increasing, and $\tilde{m}=NF(x)$, since
\[ \Pr(\tilde{m}\leq y)=\Pr(x\leq F^{-1}(y/N))=F(F^{-1}(y/N))=y/N.\]
Now, since we require a discrete distribution, we take $\bar{m}=\lfloor \tilde{m}\rfloor$.
We invert this by taking $\hat{m}=\lfloor F^{-1}(\bar{m})\rfloor$. Now, since $F$ is  strictly increasing,
\begin{align*}
   \hat{m} &= \lfloor F^{-1}(\bar{m}/N)\rfloor \leq F^{-1}(\tilde{m}/N)<F^{-1}(NF(m+1)/N)=m+1  \\
   \hat{m} &= \lfloor F^{-1}(\bar{m}/N)\rfloor > F^{-1}((\tilde{m}-1)/N)\geq F^{-1}(NF(m-1)/N)=m-1,
\end{align*}
and so $\hat{m}=m$. Thus the transformation is uniquely invertible. Of course, this does not imply that $\hat{m}$ and $m$ will have exactly the same distribution, but we may also calculate
\begin{align*}
   \Pr(\hat{m}\leq x) &\leq \Pr(\bar{m}\leq NF(x)) < Pr(\tilde{m}\leq NF(x)+1) = F(x)+1/N, \\
    \Pr(\hat{m}\leq x) &\geq \Pr(\bar{m}< NF(x+1)) \geq Pr(\bar{m}< NF(x)) = F(x).
\end{align*}
This holds, in particular, for integers $x\in[0,M)$. Thus the total variation distance between the distributions of $\hat{m}$ and $m$ is at most $M/N$. Thus the difference between the distributions of $m$ and $\hat{m}$ will be negligible, since $N\gg M$.

This flattening allows us to satisfy the assumptions of the window one-wayness scenario above. The bit leakage in $m$ is increased, however. It is not difficult to show that it increases by approximately $\lg(mp_m/F(m))$, where $p_m$ is the frequency function $\Pr(\mathbf{m}=m)$. Thus the leakage remains near-optimal for near-uniform distributions, where $\alpha/M\leq p_m\leq \beta/M$, for some constants $\alpha,\beta>0$. In this case $\lg(mp_m/F(m))\leq \lg(\beta/\alpha)=O(1)$. There are also distributions which are far from uniform, but the ratio $mp_m/F(m)$ remains bounded. Further, suppose we have a distribution satisfying $1/m^\alpha\leq p_m\leq 1/m^\beta$, for constants $\alpha,\beta>0$ such that $0<\alpha-\beta<\nicefrac12$. Then $\lg(mp_m/F(m))<\nicefrac12\lg m$, so the leakage is less than in the scheme of~\cite{Boldyreva2009}.

This transformation also allows us to handle relatively small plaintext spaces $[0,M)$, by expanding them to a larger space $[0,N)$.

Finally, note that the flattening approach here is rather different from those in~\cite{agrawal2004ope} and~\cite{Yum2012}, though not completely unrelated.\vspace{-1ex}
\section{Algorithms of Boldyreva type}\vspace{-1ex}
\label{section:genericapprox}
We have chosen to compare our scheme with that of Boldyreva et al.~\cite{Boldyreva2009}, since it has been used in practical contexts by the academic community \cite[p.5]{Boldyreva2011}, as well as in Popa et al.'s original version of CryptDB \cite{popa2011cryptdb}, which has been used or adopted by several commercial organisations \cite{popa2014source}. However, scant computational experience with the scheme has been reported \cite{popa2011cryptdb}. Therefore, we believe it is of academic interest to report our experimental results with respect to Boldyreva et al.'s scheme. We also discuss some simpler variants which have better computational performance. These are compared computationally with our scheme in section~\ref{section:results} below. The relative security of the schemes has been discussed above.

In this section we describe generic encryption and decryption algorithms based on Boldyreva et al.'s algorithm \cite{Boldyreva2009}, which sample from any distribution and which bisect on the domain (section \ref{section:generic}). We also present an approximation of Boldyreva et al.'s algorithm which samples from the Beta distribution (section \ref{section:approx}). The approximation and generic algorithms are used in our experimental evaluation presented in section \ref{section:results}.\vspace{-2ex}

\subsection{Generic Algorithms}
\label{section:generic}
Algorithm \ref{alg:genericenc} below constructs a random order-preserving function $f:\mathcal{M}\to\mathcal{C}$, where $\mathcal{M}=[0,M], M=2^r$, and $\mathcal{C}=[1,N], N\geq 2^{2r}$, so that $c=f(m)$ is the ciphertext for $m \in \mathcal{M}$. Algorithm \ref{alg:genericenc} depends on a pseudorandom number generator, $P$, and a deterministic seed function, $S$. Likewise, Algorithm \ref{alg:genericdec} constructs the inverse function $f^{-1}:\mathcal{C}\to\mathcal{M}$ so that $m = f^{-1}(c)$.

\begin{algorithm*}[h]
\caption[Generic]{Generic Boldyreva-type Encryption Algorithm}
\label{alg:genericenc}
\begin{algorithmic}[1]
\Function{RecursiveEncrypt}{$a,b,f(a),f(b),m$}
    \State $x\gets (a+b)/2$
    \State $y\gets f(b)-f(a)$
    \State Initiate $P$ with seed $S(a,b,f(a),f(b))$
    \State Determine $z\in[0,y]$ pseudorandomly, so that $\Pr(z\notin[y/4,3y/4])$ is negligible\\
    \Comment The condition implies that $y$ cannot become smaller than $3N/4(1/4)^r= 3N/4M^2=3M/4$, with high probability.
    \State $f(x) \gets f(a)+z$
    \If{$x=m$}
        \State \Return $f(x)$
    \ElsIf{$x>m$}
        \State \Return{\Call{RecursiveEncrypt}{$a,x,f(a),f(x),m$})}
    \Else
        \State \Return{\Call{RecursiveEncrypt}{$x,b,f(x),f(b),m$}}
    \EndIf
\EndFunction
\State Initiate $P$ with a fixed seed $S_0$.
\State Choose $f(0),f(M)$ pseudorandomly so that $f(M)-f(0)>3N/4$
\State \Return{\Call{RecursiveEncrypt}{$0,M,f(0),f(M),m$}}
\end{algorithmic}
\end{algorithm*}

\begin{algorithm*}[h]
\caption[Generic]{Generic Boldyreva-type Decryption Algorithm}
\label{alg:genericdec}
\begin{algorithmic}[1]
\Function{RecursiveDecrypt}{$a,b,f(a),f(b),c$}
    \State $x\gets (a+b)/2$
    \State $y\gets f(b)-f(a)$
    \State Initiate $P$ with seed $S(a,b,f(a),f(b))$
    \State Determine $z\in[0,y]$ pseudorandomly
    \State $f(x) \gets f(a)+z$
    \If{$f(x)=c$}
        \State \Return $x$
    \ElsIf{$f(x)>c$}
        \State \Return{\Call{RecursiveDecrypt}{$a,x,f(a),f(x),c$}}
    \Else
        \State \Return{\Call{RecursiveDecrypt}{$x,b,f(x),f(b),c$}}
    \EndIf
\EndFunction
\State Initiate $P$ with a fixed seed $S_0$.
\State Choose $f(0),f(M)$ pseudorandomly so that $f(M)-f(0)>3N/4$
\State \Return{\Call{RecursiveDecrypt}{$0,M,f(0),f(M),c$}}
\end{algorithmic}
\end{algorithm*}
\subsection{An Approximation}\vspace{-0.5ex}
\label{section:approx}
We have a plaintext space,$[1,M]$, and ciphertext space, $[1,N]$. Boldyreva et al. use bijection between strictly increasing functions $[1,M]\to[1,N]$ and subsets of size $M$ from $[1,N]$, so there are $\binom{N}{M}$ such functions. There is a similar bijection between nondecreasing functions $[1,M]\to[1,N]$ and multisets of size $M$ from $[1,N]$, and there are $N^M/M!$ such functions. If we sample $n$ points from such a function $f$ at random, the probability that $f(m_1)=f(m_2)$ for any $m_1\neq m_2$ is at most $\binom{n}{2}\times 1/N<n^2/2N$. We will assume that $n\ll \sqrt{N}$, so $n^2/2N$ is negligible. Hence we can use sampling either with or without replacement, whichever is more convenient.

Suppose we have sampled such a function $f$ at points $m_1<m_2<\cdots<m_k$, and we now wish to sample $f$ at $m$, where $m_i<m<m_{i+1}$. We know $f(m_i)=c_i$, $f(m_{i+1})=c_{i+1}$, and let $f(m)=c$, so $c_i\leq c \leq c_{i+1}$.\footnote{We can have equality because we sample with replacement.}  Let $x=m-m_i$, $a=m_{i+1}-m_i-1$, $y=c-c_i$, $b=c_{i+1}-c_i+1$, so $1\leq x \leq a$ and $0\leq y\leq b$. Write $\tilde{f}(x)=f(x+m_i)-c_i$. Then, if we sample $a$ values from $[0,b]$ independently and uniformly at random, $c-c_i$ will be the $x$th smallest. Hence we may calculate, for $0\le y\leq b$,
\begin{equation}\label{ope:eq10}
\Pr\big(\tilde{f}(x)=y\big) = \dfrac{a!}{(x-1)!\,(a-x)!}\left(\dfrac{y}{b}\right)^{x-1}\dfrac{1}{b}\left(\dfrac{b-y}{b}\right)^{a-x}
\end{equation}
This is the probability that we sample one value $y$, $(x-1)$ values in $[0,y)$ and $(a-x)$ values in $(y,b]$, in any order. If $b$ is large, let $z=y/b$, and d$z=1/b$, then \eqref{ope:eq10} is approximated by a continuous distribution with, for $0\leq z\leq 1$,
\begin{equation}\label{ope:eq20}
\Pr\big(z\leq \tilde{f}(x)/b<z+\textrm{d}z\big) = \dfrac{z^{x-1}(1-z)^{a-x}}{\textrm{B}(x,a-x+1)}\,\textrm{d}z\
\end{equation}
which is the B$(x,a-x+1)$ distribution. Thus we can determine $f(m)$ by sampling from the Beta distribution to $\lg N$ bits of precision. In fact, we only need $\lg b$ bits. However, using $n\leq M\leq \sqrt{N}$,
\[
\Pr(\exists i: m_{i+1}-m_i< N^{1/3}) \,\leq\,  \frac{nN^{1/3}}{N}\,\leq\, \frac{M}{N^{2/3}}
  \,\leq\,  \frac{1}{N^{1/6}}
\]
is very small, so we will almost always need at least $\tfrac13\lg N$ bits of precision. Thus the approximation given by~\eqref{ope:eq20} remains good even when $a=1$, since it is then the uniform distribution on $[0,b]$, where $b\geq N^{1/3}$ with high probability.

When the $m_i$ arrive in random order, the problem is to encrypt them consistently without storing and sorting them. Boldyreva et al. use binary search. 
If $M=2^r$, we will always have $a=2^s$ and $x=2^{s-1}$ in~\eqref{ope:eq20}, so $a-x=x$, and~\eqref{ope:eq20} simplifies to
\begin{equation*}
\Pr\big(z\leq \tilde{f}(x)/b<z+\textrm{d}z\big) = \dfrac{z^{x-1}(1-z)^{x}}{\textrm{B}(x,x+1)}\,\textrm{d}z,
\end{equation*}
for $0\leq z\leq 1$,
This might be closely approximated by a Normal distribution if Beta sampling is too slow.\vspace{-3ex}


\section{Experimental Results}\vspace{-2ex}

\label{section:results}
To evaluate our scheme in practice, we conducted a simple experiment to pseudorandomly generate and encrypt 10,000 $\rho$-bit integers. The ciphertexts were then sorted using a customised TeraSort MapReduce (MR) algorithm \cite{omalley2008terasort}. Finally, the sorted ciphertexts were decrypted and it was verified that the plaintexts were also correctly sorted.
\begin{table*}[h]
\centering
\caption{Timings for each experimental configuration. $\rho$ denotes the bit length of the unencrypted inputs. \emph{Init} is the initialisation time for the encryption/decryption algorithm, \emph{Enc} is the mean time to encrypt a single integer, \emph{Exec} is the MR job execution time, \emph{Dec} is the mean time to decrypt a single integer.}
\label{tab:results}
\begin{tabular}{lllllll}
\toprule
Algorithm &  $\rho$ &  \multicolumn{2}{c}{Encryption} &  MR Job & \multicolumn{2}{c}{Decryption} \\
 &  & Init. (ms) & Enc. ($\mu$s) & Exec. (s) & Init. (ms) & Dec. ($\mu$s) \\
\midrule
GACD      & 7   & 50.13                & 1.51               & 63.79               & 11.62                & 1.47               \\
GACD      & 15  & 58.04                & 2.18               & 61.28               & 10.86                & 2.46               \\
GACD      & 31  & 58.66                & 2.07               & 63.02               & 12.18                & 2.59               \\
GACD      & 63  & 70.85                & 1.94               & 65.20               & 10.61                & 4.22               \\
GACD      & 127 & 91.94                & 2.38               & 61.08               & 11.10                & 6.29               \\
BCLO & 7   & 143.72               & 191.48             & 70.78               & 154.01               & 192.42             \\
BCLO & 15  & 135.04               & 74390.95           & 65.47               & 148.29               & 79255.23           \\
Beta      & 7   & 189.52               & 57.87              & 64.77               & 208.16               & 58.27              \\
Beta      & 15  & 202.64               & 124.79             & 63.70               & 218.91               & 121.53             \\
Beta      & 31  & 181.14               & 221.92             & 63.64               & 208.22               & 221.83             \\
Beta      & 63  & 176.24               & 477.23             & 66.74               & 193.03               & 466.03             \\
Uniform   & 7   & 167.66               & 42.61              & 64.64               & 182.27               & 42.92              \\
Uniform   & 15  & 166.98               & 83.40              & 66.29               & 176.14               & 82.53              \\
Uniform   & 31  & 162.11               & 179.92             & 63.89               & 176.53               & 180.52             \\
Uniform   & 63  & 156.53               & 409.13             & 63.91               & 173.57               & 412.79             \\
Uniform   & 127 & 162.17               & 1237.34            & 65.30               & 170.74               & 1232.19            \\
\bottomrule
\end{tabular}
\end{table*}

The MR algorithm was executed on a Hadoop cluster of one master node and 16 slaves. Each node was a Linux virtual machine (VM) having 1 vCPU and 2GB RAM. The VMs were hosted in a heterogeneous OpenNebula cloud. In addition, a secure Linux VM having 2 vCPUs and 8 GB RAM was used to generate/encrypt and decrypt/verify the data.\\\indent
Our implementation is pure, unoptimised Java utilising the JScience library \cite{jscience2014} arbitrary precision integer classes. It is denoted as algorithm \emph{GACD} in Table \ref{tab:results}. In addition, to provide comparison for our algorithm we have implemented Boldyreva et al.'s algorithm (referred to as \emph{BCLO}) \cite{Boldyreva2009} along with two variants of the Boldyreva et al. algorithm. These latter variants are based on our generic version of Boldyreva et al.'s algorithm (see section \ref{section:generic}). One is an approximation of Boldyreva'a algorithm which samples ciphertext values from the Beta distribution (referred to as \emph{Beta} in Table \ref{tab:results}). The derivation of this approximation is given in section \ref{section:approx}. The second samples ciphertexts from the uniform distribution (referred to as \emph{Uniform} in Table \ref{tab:results}). This variant appears in Popa et al.'s CryptDB \cite{popa2011cryptdb} source code \cite{popa2014source} as \texttt{ope-exp.cc}. The mean timings for each experimental configuration is tabulated in  Table \ref{tab:results}. The chosen values of $\rho$ for each experimental configuration are as a result of the implementations of Boldyreva et al. and the Beta distribution version of the generic Boldyreva algorithm. The Apache Commons Math \cite{commonsmath2016} implementations of the hypergeometric and Beta distributions we used only support Java signed integer and signed double precision floating point parameters respectively, which account for the configurations seen in Table \ref{tab:results}. To provide fair comparison, we have used similar configurations throughout. It should be pointed out that, for the \emph{BCLO}, \emph{Beta} and \emph{Uniform} algorithms, when $\rho=7$, this will result in only 128 possible ciphertexts, even though we have 10,000 inputs. This is because these algorithms will only encrypt each plaintext to a unique value. Such a limited ciphertext space makes these algorithms trivial to attack. Our algorithm will produce 10,000 different ciphertexts as a result of the ``noise'' term. Each ciphertext will have an effective entropy of at least 21 bits for $\rho=7$ (see section \ref{section:security}). So, our algorithm is more secure than \emph{BCLO}, \emph{Beta}, and \emph{Uniform} for low entropy inputs.\\\indent
As shown by Table \ref{tab:results}, our work compares very favourably with the other schemes. The encryption times of our algorithm outperform the next best algorithm (\emph{Uniform}) by factors of 28 ($\rho=7$) to 520 ($\rho=127$). Furthermore, the decryption times grow sublinearly in the bit length of the inputs. Compare this with the encryption and decryption times for the generic Boldyreva algorithms which, as expected, grow linearly in the bit length of the inputs. Boldyreva et al.'s version performs even worse. We believe this is down to the design of the algorithm, as stated in~\cite{Boldyreva2009}, which executes $n$ recursions where $n$ is the bit-size of the ciphertexts. We also discovered that the termination conditions of their algorithm can result in more recursions than necessary.\\\indent
It should also be noted that the size of the ciphertext generated by each algorithm seems to have minimal bearing on the MR job execution time. Table \ref{tab:results} shows that the job timings are similar regardless of algorithm.\\ \indent
Of course, it is impossible to compare the security of these systems experimentally, since this would involve simulating unknown attacks. But we have shown above that the GACD approach gives a better theoretical guarantee of security than that of~\cite{Boldyreva2011,Boldyreva2009,teranishi2014oneway}, which defines security based on a game, rather than on the conjectured hardness of a known computational problem.\vspace{-2ex}
\section{Conclusion}\vspace{-2ex}
\label{section:conc}
Our work has produced an OPE scheme based on the general approximate common divisor problem (GACDP). This appears to be the first OPE scheme to be based on a computational hardness primitive, rather than a security game. We have described and discussed the scheme, and proved its security properties, in section~\ref{section:ope}. In section~\ref{section:results} we have reported on experiments to evaluate its practical efficacy, and compare this with the scheme of ~\cite{Boldyreva2009}. Our results show that our scheme is very efficient, since there are $O(1)$ arithmetic operations for encryption and decryption. As a trade-off against the time complexity of our algorithms, our scheme produces larger ciphertexts, $\sim$ 3.67 times the number of bits of the plaintext. However, as pointed out in section \ref{section:results}, ciphertext sizes had minimal impact on the running time of the MR job used in our experiments.

With regard to our stated purpose, our experimental results show that the efficiency of our scheme makes it suitable for practical computations in the cloud.

We have noted that, like any ``true'' OPE, our scheme cannot guarantee indistinguishability under CPA~\cite{Boldyreva2009}, unlike the non-OPE protocols of Popa and others~\cite{popa2013ideal,kerschbaum2014oai}. However, with proper choice of parameters, we believe that its security is strong enough for the purpose for which it is intended: outsourcing of computation to the cloud.
\vspace{-2ex}

\printbibliography
\end{document}